\documentclass[11pt]{article}

\usepackage{graphicx}
\usepackage{times}

\usepackage{bm}
\usepackage{latexsym}
\usepackage{amsmath}
\usepackage{amssymb}
\usepackage{bm}
\usepackage{color}

\def\be{\begin{equation}}
\def\ee{\end{equation}}
\def\bea{\begin{eqnarray}}
\def\eea{\end{eqnarray}}

\newenvironment{sciabstract}{%
\begin{quote} \bf}
{\end{quote}}

\title{Photon echoes retrieved from semiconductor spins: \\ access to basis for long-term optical memories}

\author
{L.~Langer,$^{1}$ S.V.~Poltavtsev,$^{1,2}$ I.A.~Yugova,$^{2}$ M.~Salewski,$^{1}$ D.R.~Yakovlev,$^{1,3}$\\
G.~Karczewski,$^{4}$ T.~Wojtowicz,$^{4}$ I.A.~Akimov,$^{1,3}$ and M.~Bayer$^{1}$
\\
\\
\normalsize{$^{1}$Experimentelle Physik 2, Technische Universit\"at Dortmund, 44221 Dortmund, Germany}\\
\normalsize{$^{2}$Spin Optics Laboratory, Saint Petersburg State University, 198504 St. Petersburg, Russia}\\
\normalsize{$^{3}$A.F. Ioffe Physical-Technical Institute, Russian Academy of Sciences, 194021 St. Petersburg, Russia}\\
\normalsize{$^{4}$Institute of Physics, Polish Academy of Sciences, PL-02668 Warsaw, Poland}\\
\\
}

\date{}

\begin{document}


\baselineskip15pt


\maketitle


\begin{sciabstract}
The possibility to store optical information is important for
classical and quantum communication. Atoms or ions as well as color
centers in crystals offer suitable two-level systems for absorbing
incoming photons. To obtain a reliable transfer of coherence, strong
enough light-matter interaction is required, which may enforce use
of ensembles of absorbers, but has the disadvantage of unavoidable
inhomogeneities leading to fast dephasing. This obstacle can be overcome by echo techniques that
allow recovery of the information as long as the coherence is preserved.
Albeit semiconductor quantum structures appear appealing for information
storage due to the large oscillator strength of optical transitions, inhomogeneity typically is even more pronounced for them
and most importantly the optical coherence is limited to nanoseconds or
shorter. Here we show that by transferring the information to
electron spins the storage times for the optical coherence can be extended by
orders of magnitude up to the spin relaxation time. From the spin reservoir it
can be retrieved on purpose by inducing a stimulated photon echo.
We demonstrate this for an n-doped CdTe/(Cd,Mg)Te quantum well for which the storage
time thereby could be increased by more than three orders of magnitude
from the picosecond-range up to tens of nanoseconds.
\end{sciabstract}

Photon echoes are amazing optical phenomena in which resonant
excitation of a medium by short optical pulses results in a delayed
response in form of a coherent optical flash. Since their
first observation in ruby in 1964 \cite{Hartmann-ruby-64}, photon
echoes were reported for atom
vapors~\cite{Hartmann-PE-Na-vapour-79}, rare earth
crystals~\cite{Takeuchi-PE-RareEarth-74} and
semiconductors~\cite{Goebel-90,Web-Cundiff-91}. Spontaneous (two-pulse) and stimulated (three-pulse)
photon echoes were demonstrated and
used for studying the involved energy levels and the coherence
evolution of the optical
transitions~\cite{Wiersma-Science-87,Chemla-Shah-01,Samartsev-review}.
Currently there is great interest in application of photon echoes
for quantum memories~\cite{Lvovsky-09,Hamerer-10}. Photon echoes
occur in an ensemble of oscillators with an inhomogeneous
distribution of optical transitions. Such an ensemble
provides high efficiency and large bandwidth allowing one to store
multiple photons with high capacity. Current activities on photon
echoes have mainly concentrated on rare earth crystals and
atomic vapors with long storage times, that are crucial for
implementation of robust light-matter interfaces.

Already at the early stage of photon echo experiments the spin level
structure of ground and excited states was recognized to contribute to the formation of spontaneous and stimulated photon echo
signals~\cite{Lambert-Ruby-69,Hartmann-Pr-79,Wiersma-TT-79}. If
optically addressed states possess orbital and/or spin angular
momenta then the splitting of these states by a magnetic field (the
Zeeman effect) provides an additional degree of freedom for the control of photon echo through optical selection
rules~\cite{Basharov-86,Rubtsova-04,Langer-12}. Moreover, transfer
of coherence from optical to spin excitations has been suggested to
considerably extend the storage times as demonstrated for quasi-atomic systems having optical and spin coherences with comparable lifetimes~\cite{Afzelius-10}. Here we
demonstrate that a transverse magnetic field applied to a
semiconductor leads to transfer of short-lived optical coherence
into long-lived electron spin coherence. This allows one to induce
stimulated photon echoes on sub-$\mu$s time scales, exceeding
the radiative lifetime of the optical excitations by more than three
orders of magnitude. We reveal two mechanisms leading to this
extension of stimulated echo revival - coherence transfer and spin
fringes, and show that depending on the polarization configuration
of the three involved laser pulses we are able to shuffle the
optical coherence into a spin component that is directed either
along or perpendicular to the magnetic field. The spins directed
along magnetic field are free of any dephasing and are affected
little by spin relaxation, which makes this configuration highly
appealing for future applications in memory devices.

For demonstration of magnetic-field-induced stimulated photon echoes, we study a semiconductor CdTe/(Cd,Mg)Te quantum well (QW) which serves as model
system, that can be tailored for the targeted application on a detailed level by
nanotechnology. The fundamental optical excitations in
semiconductors, the excitons, possess large oscillator strength so
that resonant absorption may be achieved with close to unity
efficiency even for structure thicknesses smaller than the light
wavelength. Therefore propagation effects are not as important as in
atomic vapors and rare-earth crystals.

Ultrafast coherent spectroscopy of excitons employing laser pulses
is well established for semiconductor
nanostructures~\cite{Chemla-Shah-01}. However, for storage
applications excitons have been scarcely considered because of their
limited optical coherence time $T_2$ due to complex many body
interactions and their short radiative lifetime ($T_1
\leq 1$~ns) being the downside of the large oscillator strength. In
nanostructures such as quantum dots (QDs) the optical decoherence is
weak but still limited by radiative decay. Therefore approaches to
involve the long-lasting coherence of electron spins have been
pursued recently where most of the studies were focused on optical control of the spin~\cite{Greilich06,Yamamoto08,Awschalom08,Xu08}.
The storage and retrieval of optical coherence by its encoding in an ensemble of electron spins has not yet been addressed.

Figure 1 summarizes the experimental approach and the main results
on the optical properties of the studied QW in zero magnetic field.
We use a sequence of three excitation pulses with variable delays
$\tau_{12}$ between pulses 1 and 2 and $\tau_{23}$ between pulses 2
and 3. The duration of the pulses $\tau_p\approx 2\div3$~ps. Pulses 2 and
3 are propagating along the same direction, so that their wave
vectors are equal $\bf{k}_2=\bf{k}_3$. Both the spontaneous (PE) and
stimulated (SPE) photon echoes are then directed along
the $2\bf{k_2}-\bf{k_1}$ direction. The transients are measured by
taking the cross-correlation of the resulting four-wave mixing
signal $E_{\rm FWM}(t)$ with the reference pulse $E_{\rm ref}(t)$
using heterodyne detection as schematically shown in Fig.~1A [see
appendix A]. This allows us to
distinguish between the PE and SPE signals because of their different arrival
times at the detector.

Essential for our experiment is selection of a well-defined spin
level system, optically excitable according to clean selection
rules. That is why we did not select the neutral exciton but the
charged exciton consisting of two electrons and a hole,
which requires a resident electron population. The studied
sample comprises 20-nm thick CdTe QWs separated by
110 nm Cd$_{0.78}$Mg$_{0.22}$Te barriers. The barriers are doped
with donors which provide resident electrons for the QWs with density $n_e
\approx 10^{10}~$cm$^{-2}$. At the cryogenic temperature of 2~K in experiment
these electrons become localized in QW potential fluctuations due to well width and
composition variations~\cite{Zhukov-07}. In the photoluminescence spectrum both the neutral
($X$) and charged exciton (in short, trion $T^-$) are observed,
separated by the trion binding energy~\cite{fnt1}. In the $T^-$ ground state the two electrons form a spin
singlet state. The narrow widths of the spectral lines indicate a
high structural quality (see Fig.~1B).

To address solely the optical transition from electron to trion, the
photon energy of the laser is tuned to the lower energy flank of the
trion emission line, namely to $\hbar\omega=1.599$~eV. Thereby we also
selectively excite $T^-$ complexes with enhanced localization,
having longer optical coherence time. The laser spectrum is shown by the
dashed line in Fig.~1B. Figure~1C gives the characteristic four-wave
mixing signal taken for $\tau_{12}=23$~ps and $\tau_{23}=39$~ps,
while the delay $\tau_{\rm ref}$ of the reference pulse is scanned
relative to pulse 1. Signatures of spontaneous (PE) and stimulated
(SPE) photon echoes are clearly seen in the transients: the PE
signal appears at $\tau_{\rm ref}=2\tau_{12}$ and the SPE shows up
at $\tau_{\rm ref}=2\tau_{12}+\tau_{23}$~\cite{Wiersma-Science-87}. The
decay of the PE and SPE peak amplitudes with increasing $\tau_{12}$
and $\tau_{23}$, respectively, are shown in Fig.~1D. From these data
we evaluate decay times of $T_2=72$~ps and $T_1=45$~ps so that
$T_2 \approx 2T_1$ indicating that the loss of coherence for the
trions is mainly due to radiative decay with lifetime $\tau_r=T_1$.

The investigated electron-trion transition can be considered as
four-level system, as shown schematically in Fig.~2~\cite{fnt2}. We excite optical transitions between the doubly
degenerate electron states with spin projections $S_z=\pm 1/2$
(ground states $|1\rangle$ and $|2\rangle$) and the doubly
degenerate trion states with spin projections $J_z=\pm3/2$ (excited
states $|3\rangle$ and $|4\rangle$). The excited state spin
projections are determined by the heavy-hole total angular momentum
because the trion electrons form a singlet state ($S=0$). The selection
rules for optical excitation follow from angular momentum
conservation, i.e. $|1\rangle + \sigma^+ \rightarrow |3\rangle$ and
$|2\rangle + \sigma^- \rightarrow  |4\rangle$~\cite{fnt3}.
Here $\sigma^+$ and $\sigma^-$ denote the corresponding
circular photon polarizations. At zero magnetic field the
transitions are decoupled. In addition, the spin relaxation time of
hole $T_h$ and electron $T_e$ are long compared to the radiative
lifetime ($T_e, T_h \gg \tau_r$)~\cite{Zhukov-07}. Therefore the SPE
signal is expected to decay with $T_1=\tau_r$ when the delay $\tau_{23}$ is
increased, in full accord with the experimental data in Fig.~1D.

Application of an external magnetic field in the QW plane,
$\mathbf{B}\|x $, leads to Larmor precession of the electron spin in
the ground state. In that way the transfer of optical coherence into
long-lived electron spin coherence can be achieved and a dramatic
increase of the SPE decay time by several orders of magnitude may be
accomplished. There are two different mechanisms which contribute to
magnetic-field-induced signal (see Fig.~2). The first one (A) is
based on direct transfer of optical coherence into electron spin
coherence, and the second one (B) is due to de-synchronization of
the spectral gratings for electron and trion spins.

For simplicity, let us consider a situation when
the pulse duration $\tau_p$ is significantly shorter
than the period of Larmor precession
$T_L=2\pi/\omega_L=2\pi\hbar/g\mu_B B$, where $g$ is the electron
g-factor and $\mu_B$ is the Bohr magneton.  Under these conditions
the selection rules for optical transitions remain unchanged. For
the mechanisms (A) and (B) it is important that the Larmor
precession frequencies of the trion and electron spins are
different. This is perfectly well the case in QW structures where
the confinement along $z$ direction splits the heavy-hole and light-hole
bands and therefore the optically excited trion states do not
become coupled by the weak transverse magnetic field because the
transverse heavy-hole g-factor $g_{hh} \approx 0$~\cite{Debus-13}.
This feature simplifies analysis of the system's time evolution
because one has to account only for Larmor precession of the
electron spin, i.e. a periodic exchange between $S_y$ and $S_z$ with
frequency $\omega_L$. Also for simplicity, we consider the following
relations between the time constants representing realistically the
situation for the studied electron-trion system and corresponding as
well to the most interesting case of long-lived echoes:
\begin{eqnarray}
\label{eq:1}
& T_L \leq \tau_{12} \ll T_2, &\\
\label{eq:2}
& T_1 \ll \tau_{23} \ll T_e ~~ {\rm and} ~~ T_1 \ll T_h. &
\end{eqnarray}
The first relation requires fast Larmor spin precession and
conservation of optical coherence before arrival of pulse 2. The
second relation limits our consideration to arrival times of pulse 3
after trion recombination, i.e. the system is in the ground state at
$t=\tau_{23}$ and all required information is stored in the electron
spin. Here we remind that $\tau_r \approx T_1\approx T_2/2$. If
relations 1 and 2 hold the solutions of the Lindblad equation of
motion for the ($4\times4$) density matrix $\rho_{ij}$, which describes the
four-level electron-trion system, can be written in a compact form.
For the electron spin components containing the required terms with
$\exp(-i\omega_0\tau_{12})$ time evolution (see appendix B) we obtain:
\begin{eqnarray}
\label{eq:3}
S_x & = & \frac{\rho_{12} + \rho_{21}^*}{2} \propto K \Sigma \sin{\left(\frac{\omega_L\tau_{12}}{2}\right)} \exp{\left(-\frac{\tau_{23}}{T_1^e}\right)}, \\
\label{eq:4}
S_y & = & \frac{\rho_{12}-\rho_{21}^*}{2i} \propto  K \Delta \sin{\left(\frac{\omega_L\tau_{12}}{2} + \omega_L\tau_{23} \right)} \exp{\left(-\frac{\tau_{23}}{T_2^e}\right)}, \\
\label{eq:5}
S_z & = &  \frac{\rho_{11}-\rho_{22}}{2} \propto  - K \Delta \cos{\left(\frac{\omega_L\tau_{12}}{2} + \omega_L\tau_{23} \right)} \exp{\left(-\frac{\tau_{23}}{T_2^e}\right)}.
\end{eqnarray}
Here $K=\exp[-i(\omega_0\tau_{12}-{\bf k_1r} + {\bf
k_2r})]\exp(-\tau_{12}/T_2)+c.c.$ is the term that carries the
information on the optical phase,
$\Delta=\theta_{1+}\theta_{2+}-\theta_{1-}\theta_{2-}$ and
$\Sigma=\theta_{1+}\theta_{2+}+\theta_{1-}\theta_{2-}$ account for
$\sigma^\pm$  light polarization of the excitation pulse $n$ with
pulse area $\theta_{n\pm} \ll 1$. $T_1^e$ and $T_2^e$ correspond to
the longitudinal and transverse electron spin relaxation times, respectively.
$\omega_0$ is the trion resonance frequency. Here we assume that
initially, before the arrival of pulse 1 ($t<0$), the electron spins
are unpolarized, i.e., $\rho_{11}=\rho_{22}=1/2$, while all other
elements of the density matrix vanish. From equations 3-5 it follows
that at $B>0$ all spin components are finite with magnitudes depending
critically on the polarization of the exciting pulses.

Qualitatively the magnetic-field-induced SPE evolution is easy to
follow for a circular polarized pulse sequence as shown in Fig.~2.
At $t=0$ pulse 1 creates a coherent superposition between the
states $|1\rangle$ and $|3\rangle$. This is an optical coherence
associated with the $\rho_{13}$ element of the density matrix. Due
to inhomogeneous broadening of optical transitions, this
coherence $\rho_{13}$ disappears due to dephasing. Each dipole in
the ensemble with a particular optical frequency $\omega_0$ acquires
an additional phase $\phi(t) = (\omega-\omega_0) t = \delta\omega_0
t$ before arrival of the second pulse at $t=\tau_{12}$ (see term $K$
in Eqs.~3-5). This is indicated by a set of arrows with different
lengths, symbolizing the phase distribution of the dipoles with
different frequencies.

The first mechanism (A) (see Fig.~2A) is most efficient when the
second pulse arrives at $\tau_{12}=(2m+1)\pi/\omega_L$, where $m$ is
an integer, i.e. the optical coherence $\rho_{13}$ is shuffled into
a $\rho_{23}$ coherence between the optically inaccessible states
$|2\rangle$ and $|3\rangle$. Pulse 2 transfers this coherence into a
superposition of states $|1\rangle$ and $|2\rangle$, corresponding
to spin coherence $\rho_{12}=S_x-iS_y$ (see Eqs.~3 and 4). There the
coherence is frozen in the ground state without any further optical
dephasing and can survive for much longer times than the
zero-field coherence, even after radiative trion recombination
$\tau_r$. Note, however, that the Larmor precession of the electron
spin is continuing then. Therefore dephasing of the $S_y$ component
may occur, while $S_x$ remains constant as it is directed along the
magnetic field (see Eqs.~3 and 4). Finally, pulse 3 retrieves the
coherence $\rho_{12}$ by converting it back into the optical
frequency domain and starting the rephasing process. Again, the
rephasing will be most efficient if
$\rho_{12}(t=\tau_{12}+\tau_{23}) = \rho_{12}^*(t=\tau_{12})$, i.e.
for $\tau_{23}=(2l+1)\pi / \omega_L$, where $l$ is an integer.

The second mechanism (B) (see Fig.~2B) can be considered as an
incoherent one. In contrast to the first one it relies on population
rather than coherence. Here, the accumulated phase of each dipole
$\phi$ is projected by pulse 2 into population interference fringes
$\rho_{11} \propto \sin^2(\delta\omega_0\tau_{12}/2)$ and $\rho_{33}
\propto \cos^2(\delta\omega_0\tau_{12}/2)$, i.e. into spectral
population gratings in the excited and ground states with opposite
phase~\cite{Wiersma-Science-87}. They are equivalent to spectral
spin gratings for the electrons $S_z=(\rho_{11}-\rho_{22})/2$ and
the trions $J_z=(\rho_{33}-\rho_{44})/2$. In contrast to the previous
mechanism the interference fringes have the highest contrast when
the second pulse arrives after an integer number of electron spin
revolutions $t=\tau_{12}=2m\pi/\omega_L$ (see Eq.~5). For $t>\tau_{12}$ the
Larmor precession of the electron spin de-synchronizes the spin
gratings in ground and excited state. Therefore even after trion
recombination the electron spin grating does not disappear. The spin
fringes $S_z(\delta\omega_0)$ and $J_z(\delta\omega_0)$ are
schematically shown in Fig.~2C,D for two different times: the time
of their creation $t=\tau_{12}$ and the time after trion
recombination before arrival of pulse 3. Accordingly a long-lived
electron spin grating is present, allowing one to retrieve the phase
information $\phi$ and observe the SPE pulse with maximum signal at
$\tau_{23}=l\pi/\omega_L$.

In practice, the initial condition of zero electron spin
polarization is hard to match for circularly polarized pulse
sequences. This is because $\sigma^+$ excitation induces a
macroscopic spin polarization. The electron spin relaxation $T_2^e\sim 30$~ns
is longer than the pulse repetition period $T_R=13.2$~ns in our
experiment and $\mathbf{S}(t=0)\neq0$~\cite{Zhukov-07}. Moreover, for arbitrary
$\tau_{12}$, both mechanisms are present and therefore all spin components contribute to the SPE signal.
Therefore circularly polarized pulses are not optimal for
demonstration of magnetic-field-induced SPE. From Eqs. 3-5 it
follows that a linearly polarized pulse sequence is much more
attractive. For linearly $H=(\sigma^+ + \sigma^-)/\sqrt2$ or
$V=(\sigma^+ -\sigma^-)/\sqrt2$ co-polarized pulses, $\Delta=0$ and
$\Sigma = 2\theta_1\theta_2$ so that only the dephasing-free $S_x$
component is involved. Here we exploit only the first mechanism (A) of coherence transfer. If the pulses 1 and 2 are cross-polarized the
opposite situation with $\Delta = 2\theta_1\theta_2$ and $\Sigma=0$
is obtained. Here, the $S_y$ and $S_z$ components contribute. The
corresponding SPE amplitudes are given by
\begin{eqnarray}
P_{\rm HHHH} &\propto& \sin^2\left(\frac{\omega_L\tau_{12}}{2}\right) \exp{\left(-\frac{\tau_{23}}{T_1^e}\right)}\\
\label{eq:6}
P_{\rm HVVH} &\propto& \cos\left[\omega_L(\tau_{12}+\tau_{23}) \right] \exp{\left(-\frac{\tau_{23}}{T_2^e}\right)},
\label{eq:7}
\end{eqnarray}
where the polarization sequence is denoted by the subscript ABCD with A, B, C
corresponding to the polarizations of pulses 1, 2, 3, respectively,
and D is the polarization of the resulting SPE pulse (see appendix B). Note
that the SPE echo in HVVH configuration oscillates with the Larmor
precession frequency in dependence of $\tau_{12}+\tau_{23}$, while in
HHHH configuration it varies only with the $\tau_{12}$ delay time.
Also the decay of the signals is different. For HVVH the decay
occurs according to the transverse spin relaxation time $T^e_2$ and
additional dephasing due to the electron g-factor inhomogeneity $\Delta g$ may play a role. In HHHH the signal decays with the
longitudinal spin relaxation time $T^e_1$.

The experimental data for the SPE amplitude as function of delay
time $\tau_{23}$ and magnetic field $B$, measured in the HHHH and
HVVH polarization configurations, are summarized in Fig.~3. The
delay time between pulses 1 and 2 is set to $\tau_{12}=27$~ps, which
corresponds to $T_L/\tau_{12}=0.40$ at $B=0.7$~T and
$T_L/\tau_{12}=0.086$ at $B=0.15$~T. For comparison the SPE decay
measured at $B=0$ is shown in Fig.~3A. The magnetic field dependence of the SPE amplitude in Fig.~3B and 3D is measured for $\tau_{23}=1.27$~ns which is significantly longer than the radiative lifetime $T_1$. Note that the experimentally measured signal
corresponds to the absolute value of amplitude $|P|$. In full accord with our
expectations we observe a long-lived SPE signal when applying the
magnetic field. The
theoretical curves are in good agreement with the data.
The calculations take into account the full dynamics of the system,
including the excited states, and therefore reproduce also the short
decay component of the SPE due to trion recombination as well as the signals
at low magnetic fields with $T_L>\tau_{12}$ (see appendix B).

In case of HHHH no oscillations appear when $\tau_{23}$ is varied (see Fig.~3A),
which is in line with Eq.~6. Particularly fascinating is the
observation of SPE signal at negative delays ($\tau_{23}\sim-300$~ps)
at a 75\% level of the SPE amplitude at positive delays
($\tau_{23}\sim 300$~ps), see Fig.~3A. This means that within the
pulse repetition period of 13~ns the SPE amplitude reduces by 25\%
only. From these data we estimate $T_1^e\sim 50$~ns, which allows
us to observe SPE signals in magnetic field on a sub-$\mu$s
timescale. The short dynamics at negative delays $-200~{\rm ps} <
\tau_{23} < 0$ is due to excitation of trions by pulse 3, which
influences the initial conditions at $t=0$. Note that for the
linearly polarized pulse sequence no macroscopic spin polarization
becomes involved in the PE experiment due to inhomogeneous
broadening of the optical transitions. Therefore we obtain
information about the intrinsic longitudinal spin relaxation of the
electron spin. The experimental data demonstrate that transfer of
coherence to $S_x$ is feasible and attractive because this spin
component being parallel to $\bf B$ is robust against relaxation and is not sensitive to
dephasing. In contrast, the oscillatory signal in the HVVH
polarization configuration decays much faster due to electron
g-factor inhomogeneity and consequently no long-lived SPE signal is
observed at negative delays (see Fig.~3C). The reduction of SPE
signal at $B>0.3$~T in Fig.~3D is also related to this fact. From fitting
the data we evaluate $\Delta g = 0.018$.

In conclusion, we have demonstrated magnetic-field-induced
long-lived stimulated photon echoes in the electron-trion system. By
a proper choice of polarization pulse sequence optical coherence can
be transferred into spin directed along magnetic field. Although no
metastable states are involved, due to the long-lived electron spin
coherence, the timescale of echo stimulation can be extended by more
than 3 orders of magnitude over the optical coherence time in the QW
system. Note that the electron-trion energy level structure is identical in
QWs and self-assembled quantum dots (QDs). We used QWs
for demonstration purpose because the trion transitions are well
isolated spectrally. As a downside, we had to keep the excitation
power low in order to suppress many-body effects. In case of singly
charged QD structures $\pi/2$ and $\pi$ pulses can be efficiently
used for coherent manipulation~\cite{QDs-Rabi}. In addition the
longitudinal spin relaxation times in, e.g., (In,Ga)As QDs may be as long as
0.1~s~\cite{Finley04}. Further exploiting hyperfine interaction between electrons and nuclei might enable
storage time of seconds or longer~\cite{Oulton07}. Therefore our findings open a new avenue for
realization of optical memories in semiconductor nanostructures.

\vskip 2cm
\noindent{\bf Acknowledgements}

The Dortmund team would like to acknowledge financial support of this work by the Deutsche Forschungsgemeinschaft, the Bundesministeriuum f\"ur Bildung und Forschung (project Q.com-H). The project "SPANGL4Q" acknowledges the financial support of the Future and Emerging Technologies (FET) programme within the Seventh Framework Programme for Research of the European Commission, under FET-Open grant number: FP7-284743.  S.V.P. and I.A.Yu. acknowledge partial financial support from the Russian Ministry of Science and Education (contract No.11.G34.31.0067). The research in Poland was partially supported by the National Science Center (Poland) under the Grants DEC-2012/06/A/ST3/00247 and DEC-2013/ST3/229756.

\vskip 1cm

\newpage
\appendix
\section{Methods}

The semiconductor quantum well (QW) structure was grown by molecular-beam epitaxy. It comprises 5 electronically decoupled 20~nm thick CdTe QWs embedded in 110~nm Cd$_{0.78}$Mg$_{0.22}$Te barriers. The barriers are doped by iodine donors, which provide the QW layers with conduction band electrons of low density $n_e \sim 10^{10}$~cm$^{-2}$~\cite{Zhukov-07}. The sample was mounted into a liquid He bath cryostat at a temperature of 2 K. Magnetic fields up to 0.7~T were applied  in Voigt geometry using an electromagnet. The direction of the magnetic field was parallel to the QW plane (${\bf B} \| x$).

We used tunable self mode-locked Ti:Sa laser as source of the optical pulses with durations of $2\div3$~ps at a repetition rate of 75.75~MHz (repetition period $T_R=13.2$~ns). The laser was split into four beams. Three of them were used for the three pulse sequence required for stimulating the photon echo. The fourth beam was used as reference pulse in the heterodyne detection. The delay between all four pulses could be scanned by reflectors mounted on mechanical translation stages. The three-pulse four-wave mixing (FWM) experiment was performed in reflection geometry. Pulse 1 with wavevector $\bf k_1$ hit the sample under an incidence angle of about 7$^\circ$. Pulses 2 and 3, both traveled along the same direction ($\bf k_2=k_3$) different from that of the first beam, hit the QW structure under an incidence angle of about $6^\circ$.  The beams were focused onto the sample in a spot of about $200~\mu$m in diameter. The intensities of each pulse were selected such as to remain in the linear excitation regime for each of the beams (pulse energy around 10-100~nJ/cm$^{2}$). The FWM signal was collected along the $2\mathbf{k_2}-\mathbf{k_1}$ direction. We used interferometric heterodyne detection where the FWM signal and the reference beam are overlapped at a balanced detector \cite{heterodyne}. The optical frequencies of pulse 1 and reference pulse were shifted by 40~MHz and 41~MHz with acousto-optical modulators. The resulting interference signal at the photodiode was filtered by a high frequency lock-in amplifier selecting $|2\omega_2-\omega_1-\omega_{\mathrm{ref}}|=1$~MHz. This provided a high sensitivity measurement of the absolute value of the FWM electric field amplitude in real time when scanning the reference pulse delay time $\tau_{\rm ref}$, which was taken relative to the pulse 1 time arrival. The polarization of the first and the second pulses, as well as the detection polarization, were controlled with retardation plates in conjunction with polarizers.

\section{Theoretical description of magnetic-field-induced SPE}
\subsection{Stimulated photon echo in magnetic field}

Let us consider optical excitation of the negatively charged trion by a short laser pulse with frequency $\omega$ close to the trion resonance frequency $\omega_0$.
The incident electromagnetic field induces optical transitions between the electron state and the trion state creating a coherent superposition of these states. In accordance with the selection rules, $\sigma^+$ circularly polarized light creates a superposition of the $+1/2$ electron and $+3/2$ trion states, while $\sigma^-$ polarized light creates a superposition of the $-1/2$ electron and $-3/2$ trion states. In order to describe these superpositions and the resulting dynamics in a magnetic field we use a 4x4 time-dependent density matrix,
comprising the two electron spin projections ($\pm 1/2$) (index 1 and 2) and the two hole spin projections ($\pm 3/2$) (index 3 and 4).

The temporal evolution of the density matrix is described by the Lindblad equation:
\begin{equation}
\label{eq:eq1}
\dot{\rho}=-\frac{\mathrm i}{\hbar}[\hat{H},\rho]+\Gamma.
\end{equation}
Here $\hat{H}$ is the Hamiltonian of the system and $\Gamma$
describes relaxation processes phenomenologically. In our case the
Hamiltonian contains three contributions:
$\hat{H}=\hat{H}_0+\hat{H}_B+\hat{V}$, where $\hat{H}_0$ is the
Hamiltonian of the unperturbed spin system, $\hat{H}_B$ gives the
interaction with magnetic field and $\hat{V}$ describes the
interaction with light. In the calculations we use the short pulse
approximation assuming that the pulse duration is significantly
shorter than the trion lifetime, the decoherence times and the
electron spin precession period in transverse magnetic field.
This assumption is justified for our experimental conditions.
Under these circumstances, we can separate and consider consistently
the interaction of the electron-trion system with light and its
dynamics in magnetic field.

\subsection{Electron-trion system under action of short light pulse}

The interaction with the electromagnetic wave in the electric-dipole approximation is described
by the Hamiltonian:
\begin{equation}
 \label{eq:eq2}
\hat V(t) = -\int [\hat d_+(\bm r) E_{\sigma^+}(\bm r,t) +
\hat d_-(\bm r)E_{\sigma^-}(\bm r,t)] \mathrm d^3 r\:,
\end{equation}
where  $\hat d_\pm(\bm r)$ are the circularly polarized components
of the dipole moment density operator, and $E_{\sigma ^\pm}(\bm
r,t)$ are the correspondingly polarized components of the electric
field of a quasi-monochromatic electromagnetic wave. The electric
field of this wave is given by
\begin{equation}
\bm E(\bm r, t) = E_{\sigma^+}(\bm r,t) \bm o_+  + E_{\sigma^-}(\bm
r,t)\bm o_- + {\rm c.c.}\:, \label{eq:eq3}
\end{equation}
where $\bm o_\pm$ are the circularly polarized unit vectors that are
related to the unit vectors ${\bm o}_x \parallel x$ and ${\bm o}_y
\parallel y$ through $\bm o_\pm = (\bm o_x \pm \mathrm i \bm
o_y)/\sqrt{2}$. Here the components $E_{\sigma ^+}$ and
$E_{\sigma^-}$  contain temporal phase factors $ \mathrm e^{-\mathrm
i \omega t}$.

The strength of the light interaction with the electron-trion system
is characterized  by the corresponding transition matrix element of
the operators $\hat{d}_{\pm}({\bm r})$ calculated with the wave
functions of the valence band, $|\pm 3/2\rangle$, and the conduction
band, $|\pm 1/2\rangle$: \cite{ivchenko05a}
\begin{equation}
\label{dpm}
\mathsf d(\bm r) = \langle 1/2 |\hat d_- (\bm r)|3/2\rangle =
\langle - 1/2 |\hat d_+ (\bm r)|-3/2 \rangle.
\end{equation}

We assume, that the trion recombination time is considerably shorter
than the laser repetition period. Therefore, before the first pulse
only elements of the density matrix describing the electron are
unequal zero. We also assume, that these are only the populations
$\rho_{11}$ and $\rho_{22}$.
Thus, the initial conditions are:
\begin{align}
\label{initial-cond}
\rho (0) &=
\begin{pmatrix}
\rho_{11}(0) & 0  &0   &0   \\
0 &\rho_{22}(0)& 0&0\\
0&0&0&0\\
0&0&0&0
\end{pmatrix}.
\end{align}

The Hamiltonian $\hat{H}=\hat{H}_0+\hat{V}$ in our basis is given by:
\begin{align}
\frac{1}{2}
\begin{pmatrix}
0 & 0  &f_+^* \mathrm e^{\mathrm i\omega t}\hbar   &0   \\
0 &0& 0&f_-^* \mathrm e^{\mathrm i\omega t}\hbar  \\
f_+ \mathrm e^{-\mathrm i\omega t}\hbar &0&2\hbar \omega_0 &0\\
0&f_- \mathrm e^{-\mathrm i\omega t} \hbar &0&2\hbar \omega_0
\end{pmatrix}.
\end{align}
Here $f_{\pm}(t)$ is proportional to the
smooth envelopes of the circular polarized components $\sigma^+$ and
$\sigma^-$ of the excitation pulse, given by
\[
f_{\pm}(t) = -\frac{2\mathrm e^{\mathrm i \omega t}}{\hbar}\int \mathsf d(\bm r)
E_{\sigma_{\pm}}(\bm r,t)\mathrm d^3 r\:.
\]
For simplicity we hereafter consider pulses with rectangular shape.
This means that the pulse areas for the $\sigma^{\pm}$ polarized components are equal to $f_{\pm}t_p$. In the experiment we investigate localized electrons  and trions. We suppose, that the light wavelength is much larger than the length scale of localization. This allows one to extract the  electric field $E_{\sigma \pm}$ from the integral and write the pulse areas in the form $f_{\pm}t_p=\theta_{\pm} \mathrm e^{\mathrm i {\bf kr}}$, where $r$ is the position of the localized resident electron \cite{yugova09}. For calculations of the light-induced optical polarization and, therefore, the photon echo amplitude, we have to integrated the final expressions over the electron positions \cite{yugova09}. In experiment we operate in the linear regime of optical excitation of the QW, so that $|f_{\pm}t_p| \ll 1$ \cite{rsa_vs_ml}.

With these assumptions, the solution of the von Neumann
equation $\mathrm i\hbar \dot{\rho}=[\hat{H}_0+\hat{V},\rho]$ after
the pulse action gives:
\begin{subequations}
\label{rho_light}
\bea
\rho_{11}^a = \rho_{11}^b +\frac{\mathrm it_p}{2}(f_+ \rho_{13}^b - f_+^*\rho_{31}^b ) + (\rho_{33}^b - \rho_{11}^b)\frac{|f_+t_p|^2}{4},\nonumber\\
\rho_{33}^a = \rho_{33}^b -\frac{\mathrm it_p}{2}(f_+\rho_{13}^b - f_+^*\rho_{31}^b ) - (\rho_{33}^b - \rho_{11}^b)\frac{|f_+t_p|^2}{4},\nonumber\\
\rho_{22}^a = \rho_{22}^b +\frac{\mathrm it_p}{2}(f_- \rho_{24}^b - f_-^{*}\rho_{42}^b ) + (\rho_{44}^b - \rho_{22}^b)\frac{|f_-t_p|^2}{4},\nonumber\\
\rho_{44}^a = \rho_{44}^b -\frac{\mathrm it_p}{2}(f_-\rho_{24}^b - f_-^{*} \rho_{42}^b ) - (\rho_{44}^b - \rho_{22}^b)\frac{|f_-t_p|^2}{4},\nonumber\\
\eea
\bea
\rho_{13}^a = \mathrm e^{\mathrm i\omega t_p}[\rho_{13}^b -\frac{\mathrm if_+^{*}t_p}{2} (\rho_{33}^b - \rho_{11}^b)+ \frac{f_+^{*}t_p^2}{4}(f_+ \rho_{13}^b + f_+^{*}\rho_{31}^b)],\nonumber\\
\rho_{24}^a = \mathrm e^{\mathrm i\omega t_p}[\rho_{24}^b -\frac{\mathrm if_-^{*}t_p}{2}(\rho_{44}^b - \rho_{22}^b)+ \frac{f_-^{*}t_p^2}{4}(f_-\rho_{24}^b + f_-^{*} \rho_{42}^b )],\nonumber\\
\eea
\bea
\rho_{14}^a = \mathrm e^{\mathrm i\omega t_p}[\rho_{14}^b + \rho_{32}^b \frac{f_+^{*}f_-^{*}t_p^2}{4} +\frac{\mathrm if_-^{*}t_p}{2}\rho_{12}^b - \frac{\mathrm if_+^{*}t_p}{2}\rho_{34}^b],\nonumber\\
\rho_{32}^a = \mathrm e^{-\mathrm i\omega t_p}[\rho_{32}^b + \rho_{14}^b \frac{f_+ f_-t_p^2}{4} +\frac{\mathrm if_-t_p}{2}\rho_{34}^b - \frac{\mathrm if_+t_p}{2} \rho_{12}^b],\nonumber\\
\eea
\bea
\rho_{12}^a =\rho_{12}^b + \rho_{34}^b \frac{f_+^{*}f_-t_p^2}{4} +\frac{\mathrm if_-t_p}{2}\rho_{14}^b - \frac{\mathrm if_+^{*}t_p}{2}\rho_{32}^b,\nonumber\\
\rho_{34}^a = \rho_{34}^b + \rho_{12}^b \frac{f_+ f_-^{*}t_p^2}{4} +\frac{\mathrm if_-^{*}t_p}{2} \rho_{32}^b  - \frac{\mathrm if_+t_p}{2}\rho_{14}^b
\eea
\end{subequations}
The superscripts `b' and `a' denote the matrix elements before and after pulse arrival, respectively.

\subsection{Precession and relaxation in magnetic field}

Next, we have to consider the dynamics in a transverse magnetic
field. The magnetic field is applied perpendicular to the
propagation direction of the incident light and to the structure
growth axis. The corresponding Hamiltonian is:
\begin{align}
\frac{1}{2}
\begin{pmatrix}
0& \hbar \omega_L &0   &0   \\
\hbar \omega_L & 0&0&0 \\
0&0&2\hbar \omega_0 &\hbar \omega_L^T\\
0&0&\hbar \omega_L^T&2\hbar \omega_0
\end{pmatrix},
\end{align}
where $\omega_L$ and $\omega_L^T$ are the electron and trion Larmor
precession frequencies. For simplicity we neglect the trion spin
precession in magnetic field, i.e. $\omega_L^T=0$, which is
justified by the hole g-factor being close to zero.

Relaxation processes $\Gamma$ are taken into account in the
following way:
\begin{align}
\begin{pmatrix}
-\frac{\rho_{11}-\rho_{22}}{2T_{z}^e}+\frac{\rho_{33}}{\tau_r}  &  -\frac{\rho_{12}}{2T_{s}^e}-\frac{\rho_{21}}{2T_{s1}^e}&-\frac{\rho_{13}}{T_2} &-\frac{\rho_{14}}{T_2}   \\
-\frac{\rho_{12}}{2T_{s1}^e}-\frac{\rho_{21}}{2T_{s}^e}& -\frac{\rho_{22}-\rho_{11}}{2T_{z}^e}+\frac{\rho_{44}}{\tau_r}&-\frac{\rho_{23}}{T_2}&-\frac{\rho_{24}}{T_2} \\
-\frac{\rho_{31}}{T_2}&-\frac{\rho_{32}}{T_2}&-\frac{\rho_{33}-\rho_{44}}{2T_{s}^h}-\frac{\rho_{33}}{\tau_r} &-\frac{\rho_{34}}{T_{s}^h}-\frac{\rho_{34}}{\tau_r}\\
-\frac{\rho_{41}}{T_2}&-\frac{\rho_{42}}{T_2}&-\frac{\rho_{43}}{T_{s}^h}-\frac{\rho_{43}}{\tau_r}&-\frac{\rho_{44}-\rho_{33}}{2T_{s}^h}-\frac{\rho_{44}}{\tau_r}
\end{pmatrix},
\end{align}

Here $T_2$ is the decay time of the optical coherence and $\tau_r$
is the trion recombination time.
$\frac{1}{T_{s}^e}=\frac{1}{T_{x}^e}+\frac{1}{T_{y}^e}$, $\frac{1}{T_{s1}^e}=\frac{1}{T_{x}^e}-\frac{1}{T_{y}^e}$.
$T_{x,y,z}^e$ are the electron spin
relaxation times. Because the magnetic field points along the $x$
axis, we assume $T_x^e \equiv T_1^e$ (longitudinal spin relaxation time) and $T_z^e = T_y^e \equiv
T_2^e$ (transverse spin relaxation time). For the trion spin relaxation one can write down similar
decay terms, but because of $\omega_L^T = 0$ we introduce only one
the spin relaxation time $T_{s}^h$.

First, let us consider the dynamics of the non-diagonal terms of the
density matrix in magnetic field. After pulse action these elements
are: \bea \label{rhoB}
\rho_{13} (t)= [\rho_{13}^a\cos(\omega_L(t-t_0)/2)-\mathrm i \rho_{23}^a\sin(\omega_L(t-t_0)/2)]\mathrm e^{(t-t_0)(\mathrm i\omega_0-1/T_2)},\nonumber\\
\rho_{23}(t)= [\rho_{23}^a\cos(\omega_L(t-t_0)/2)-\mathrm i \rho_{13}^a\sin(\omega_L(t-t_0)/2)]\mathrm e^{(t-t_0)(\mathrm i\omega_0-1/T_2)},\nonumber\\
\rho_{24} (t)= [\rho_{24}^a\cos(\omega_L(t-t_0)/2)-\mathrm i \rho_{14}^a\sin(\omega_L(t-t_0)/2)]\mathrm e^{(t-t_0)(\mathrm i\omega_0-1/T_2)},\nonumber\\
\rho_{14}(t)= [\rho_{14}^a\cos(\omega_L(t-t_0)/2)-\mathrm i \rho_{24}^a\sin(\omega_L(t-t_0)/2)]\mathrm e^{(t-t_0)(\mathrm i\omega_0-1/T_2)}.\nonumber\\
\eea Here $t_0$ is the time after end of the pulse action.

It is convenient to describe the evolution of the other elements of
the density matrix through the electron and trion spin dynamics.
Obviously the following relations hold:
\bea
&&S_z=(\rho_{11}-\rho_{22})/2, \qquad S_y=\mathrm
i(\rho_{12}-\rho_{21})/2, \qquad
S_x=(\rho_{12}+\rho_{21})/2,\nonumber\\
&&J_z=(\rho_{33}-\rho_{44})/2, \qquad
J_y=\mathrm i(\rho_{34}-\rho_{43})/2, \qquad
J_x=(\rho_{34}+\rho_{43})/2,\nonumber\\
&&n_e=(\rho_{11}+\rho_{22})/2, \qquad
n_T=(\rho_{33}+\rho_{44})/2,\nonumber\\
\eea where $S_{x,y,z}$ and $J_{x,y,z}$ are the components of the
electron and trion spin polarization. $n_e$ and $n_T$ are the
populations of the electron and trion states.

After the excitation pulse the spin dynamics of the trion in
magnetic field is given by\cite{rsa_vs_ml}:
\bea \label{JB}
J_z(t)=J_z^a\mathrm e^{-(t-t_0)/\tau_T},\qquad J_y(t)=J_y^a\mathrm
e^{-(t-t_0)/\tau_T},\qquad
J_x(t)=J_x^a\mathrm e^{-(t-t_0)/\tau_T},\nonumber.\\
\eea Here $\tau_T$ is the trion spin lifetime, $1/\tau_T \equiv
1/T_s^h+1/\tau_r$.

The electron spin components after the pulse are: \bea \label{SB}
S_z(t) &=& \mathrm e^{-(t - t_0)/T_2^e}\left[(S_z^a+\xi_1 J_z^a)\cos(\omega_L(t-t_0))+(S_y^a+\xi_2 J_z^a)\sin(\omega_L(t-t_0)) \right]- J_z^a\xi_1 \mathrm e^{-(t - t_0)/\tau_T}\nonumber\\
S_y(t) &=& \mathrm e^{-(t - t_0)/T_2^e}\left[-(S_z^a+\xi_1 J_z^a)\sin(\omega_L(t-t_0))+(S_y^a+\xi_2 J_z^a)\cos(\omega_L(t-t_0)) \right]- J_z^a\xi_2 \mathrm e^{-(t - t_0)/\tau_T}\nonumber\\
S_x(t)& =& S_x^a\mathrm e^{-(t - t_0)/T_1^e}.
    \eea
Here the superscript $a$ denotes the spin components at time $t_0$,
when the excitation pulse has passed.
\begin{equation}
\label{xichi}
\xi_1+ \mathrm i \xi_2 = \frac{1}{\tau_r(\gamma- \mathrm i \omega)},
\end{equation}
and $\gamma = 1/\tau_T - 1/T_2^e>0$. The populations $n_e$ and $n_T$
are given by: \bea n_T(t)&=&n_T^a\mathrm e^{-(t - t_0)/\tau_r},\\
\nonumber n_e(t)&=&n_e^a+n_T^a(1-\mathrm e^{-(t - t_0)/\tau_r}).
\eea

Equations~\eqref{rho_light} as well as \eqref{JB}, \eqref{SB},
\eqref{rhoB} are basis for the following calculations and
discussion.

\subsection{Optical polarization after the third pulse}

In our experiment we measure the amplitude of the electromagnetic
wave propagating along the direction $2 \bf{k}_2-{\bf k}_1$ at delay $\tau_{12}$
after the third pulse. This amplitude is determined by the optical
polarization created in the sample by all three pulses. The
polarization in turn is proportional to the corresponding elements
of the density matrix, $\rho_{13}$, $\rho_{24}$ and c.c. averaged
over the ensemble of excited electron-trion systems. At $\tau_{12}$
after the third pulse, $\rho_{13}$ and $\rho_{24}$ for the
individual systems are given by:
\bea \label{rho13}
\rho_{13} (3t_p+2\tau_{12}+\tau_{23})= [\rho_{13}^{a3}\cos(\omega_L\tau_{12}/2)-\mathrm i \rho_{23}^{a3}\sin(\omega_L\tau_{12}/2)]\mathrm e^{\tau_{12}(\mathrm i\omega_0-1/T_2)},\nonumber\\
\rho_{24} (3t_p+2\tau_{12}+\tau_{23})=
[\rho_{24}^{a3}\cos(\omega_L\tau_{12}/2)-\mathrm i
\rho_{14}^{a3}\sin(\omega_L\tau_{12}/2)]\mathrm e^{\tau_{12}(\mathrm
i\omega_0-1/T_2)}. \eea To obtain the macroscopic polarization one
has to average these expressions for an individual electron-trion
system over the trion resonance frequency $\omega_0$. It is
obvious, that the condition for observing a stimulated photon echo
signal is excluding the factor $\mathrm e^{\mathrm
i\omega_0\tau_{12}}$ from Eq.~\eqref{rho13}. As it will be shown below, this is
possible because some contributions to $\rho_{13}^{a3}$,
$\rho_{23}^{a3}$, $\rho_{24}^{a3}$, $\rho_{14}^{a3}$ contain
$\mathrm e^{-\mathrm i\omega_0\tau_{12}}$. We turn now to
step-by-step calculations of $\rho_{13}$ and $\rho_{24}$ from
Eq.~\eqref{rho13}.

Before excitation with the first pulse only elements of the density matrix
describing the electron are unequal to zero (see Eq.~\ref{initial-cond}). We also assume, that these
non-zero elements are only populations $\rho_{11}$ and $\rho_{22}$,
$\rho_{11} = \rho_{22}=1/2$.

1. {\it The first pulse} changes electron and trion populations and
creates optical polarization, that is proportional to the elements
$\rho_{13}$, $\rho_{24}$ and c.c.
\bea \rho_{13}^{a1} = \mathrm i \theta_{1+} \mathrm e^{\mathrm
i(\omega t_p-{\bf k_1r})}/4,&
\rho_{31}^{a1} = -\mathrm i \theta_{1+} \mathrm e^{-\mathrm i(\omega t_p-{\bf k_1r})}/4,\\ \nonumber
\rho_{24}^{a1} = \mathrm i
\theta_{1-} \mathrm e^{\mathrm i(\omega t_p-{\bf k_1r})}/4,&
\rho_{42}^{a1} = -\mathrm i \theta_{1-} \mathrm e^{-\mathrm
i(\omega t_p-{\bf k_1}r)}/4.
\eea
Here the superscript `a1' indicates the values after
the first pulse action.

2. At the moment of {\it the second pulse} arrival these elements,
$\rho_{13}$, $\rho_{24}$ and c.c., (and also those arising in
magnetic field $\rho_{14}$, $\rho_{32}$ and c.c.) - in contrast to
other elements - contain the phase factor $\exp(\pm \mathrm i
\omega_0 \tau_{12})$, see Eqs.~\eqref{rho_light}. For distinctness,
the nondiagonal elements, which are proportional to $\exp(- \mathrm
i \omega_0 \tau_{12})$, are given by:
 \bea
\rho_{31}^{b2} &=& -\mathrm i \theta_{1+}\mathrm e^{-\mathrm i
(\omega t_p+\omega_0 \tau_{12}-{\bf k_1r})}
\cos(\omega_L\tau_{12}/2)\mathrm e^{-\tau_{12}/T_2}/4,\\ \nonumber
\rho_{32}^{b2} &=& \theta_{1+}\mathrm e^{-\mathrm i (\omega
t_p+\omega_0 \tau_{12}-{\bf k_1r})} \sin(\omega_L\tau_{12}/2)\mathrm
e^{-\tau_{12}/T_2}/4,\\ \nonumber \rho_{42}^{b2} &=& -\mathrm i
\theta_{1-}\mathrm e^{-\mathrm i (\omega t_p+\omega_0
\tau_{12}-{\bf k_1r})} \cos(\omega_L\tau_{12}/2)\mathrm
e^{-\tau_{12}/T_2}/4,\\ \nonumber
\rho_{41}^{b2} &=&
\theta_{1-}\mathrm e^{-\mathrm i (\omega t_p+\omega_0
\tau_{12}-{\bf k_1r})} \sin(\omega_L\tau_{12}/2)\mathrm
e^{-\tau_{12}/T_2}/4.
\eea
The superscript `b2' indicates that these
values are before the second pulse coming in.

3. The action of {\it the second pulse}, together with the already
existing optical polarization of the electron-trion system, leads to
additions to populations $\rho_{11}^{a2}$, $\rho_{22}^{a2}$,
$\rho_{33}^{a2}$, $\rho_{44}^{a2}$ and spin coherences
$\rho_{12}^{a2}$, $\rho_{34}^{a2}$, which are proportional to
$\exp(- \mathrm i \omega_0 \tau_{12})$. For convenience we rewrite
these elements of the density matrix through the components of the
electron and trion spin polarizations and populations. \bea
S_z^{a2}\sim -K \Delta \cos(\omega_L \tau_{12}/2),&S_y^{a2}\sim K
\Delta \sin(\omega_L \tau_{12}/2),&S_x^{a2}\sim -\mathrm i K \Sigma
\sin(\omega_L \tau_{12}/2),\\ \nonumber J_z^{a2}\sim  K \Delta
\cos(\omega_L \tau_{12}/2),&J_y^{a2}\sim -K \Delta_T \sin(\omega_L
\tau_{12}/2),&J_x^{a2}\sim \mathrm i K \Sigma_T \sin(\omega_L
\tau_{12}/2),\\ \nonumber n_e^{a2}\sim  -2K \Sigma \cos(\omega_L
\tau_{12}/2),&n_T^{a2}\sim 2K \Sigma \cos(\omega_L \tau_{12}/2),
\eea where \bea && K = \frac{1}{16}\mathrm e^{-\mathrm i (\omega
t_p+\omega_0 \tau_{12}-{\bf k_1r}+{\bf k_2r})}\mathrm e^{-\tau_{12}/T_2},\\
\nonumber && \Delta = \theta_{1+}\theta_{2+}-\theta_{1-}\theta_{2-},
\qquad \Sigma = \theta_{1+}\theta_{2+}+\theta_{1-}\theta_{2-},\\
\nonumber && \Delta_T =
\theta_{1+}\theta_{2-}-\theta_{1-}\theta_{2+},\qquad \Sigma_T =
\theta_{1+}\theta_{2-}+\theta_{1-}\theta_{2+}. \eea

4. Before {\it the third pulse arrival} the elements
$\rho_{11}^{b3}$, $\rho_{33}^{b3}$, $\rho_{22}^{b3}$,
$\rho_{44}^{b3}$, $\rho_{12}^{b3}$, $\rho_{34}^{b3}$, which
correspond to spin populations and spin coherences, contain only a
phase factor with optical frequency  $\exp(\pm \mathrm i \omega_0
\tau_{12})$, which arose from the second pulse action. Spin dynamics
and relaxation after the second pulse result in Larmor precession
and decay (see Eqs.~\eqref{JB},~\eqref{SB}). The other elements
contain phase factors such as $\exp(\pm \mathrm i \omega_0
\tau_{23})$, $\exp(\pm \mathrm i \omega_0 (\tau_{23}\pm
\tau_{12}))$. It should be noted, that if $\tau_{23} \gg T_2$ and
$\tau_{23} \gg \tau_r$ then only electron spin coherence and
populations can contribute to the echo signal.

The trion spin polarization shortly before the third pulse is given
by: \bea & J_z^{b3}\sim K \Delta \mathrm e^{-\tau_{23}/\tau_T}
\cos(\omega_L \tau_{12}/2),&J_y^{b3}\sim -K \Delta_T \mathrm
e^{-\tau_{23}/\tau_T} \sin(\omega_L \tau_{12}/2),\\ \nonumber &
J_x^{b3}\sim \mathrm i K \Sigma_T \mathrm e^{-\tau_{23}/\tau_T}
\sin(\omega_L \tau_{12}/2), \eea where as the electron spin
polarization is given by:
\bea & & S_z^{b3} \sim  K \Delta
\left[\mathrm e^{-\tau_{23}/T_2^e}\left[ (-1+\xi_1)\cos(\omega_L
\tau_{12}/2)\cos(\omega_L \tau_{23})+\right.\right. \\ \nonumber &&
\left. \left.  +  (\sin(\omega_L \tau_{12}/2)+\xi_2\cos(\omega_L
\tau_{12}/2))\sin(\omega_L \tau_{23})\right] -\mathrm
e^{-\tau_{23}/\tau_T}\xi_1\cos(\omega_L \tau_{12}/2)\right],\\
\nonumber && S_y^{b3} \sim  K \Delta \left[\mathrm
e^{-\tau_{23}/T_2^e}\left[ (1-\xi_1)\cos(\omega_L
\tau_{12}/2)\sin(\omega_L \tau_{23})+ \right.\right. \\ \nonumber &&
\left. \left. +(\sin(\omega_L \tau_{12}/2)+\xi_2\cos(\omega_L
\tau_{12}/2))\cos(\omega_L \tau_{23})\right] -\mathrm
e^{-\tau_{23}/\tau_T}\xi_2\cos(\omega_L \tau_{12}/2)\right],\\
\nonumber && S_x^{b3} \sim  -\mathrm i K \Sigma \mathrm
e^{-\tau_{23}/T_1^e} \sin(\omega_L \tau_{12}/2). \eea Further, the
populations $n_e$ and $n_T$ read: \bea n_e^{b3}&\sim& -2K \Sigma
\mathrm e^{-\tau_{23}/\tau_r} \cos(\omega_L \tau_{12}/2), \\
n_T^{b3}&\sim& 2K \Sigma \mathrm e^{-\tau_{23}/\tau_r} \cos(\omega_L
\tau_{12}/2) \nonumber \eea

5. The amplitude of the stimulated photon echo is proportional to
the optical polarizations terms $\rho_{13}$ and $\rho_{24}$ after {\it
the third pulse arrival} at $t=3 t_p + 2\tau_{12} + \tau_{23}$. \bea
\label{rho1324} &&\rho_{13}(3 t_p + 2\tau_{12} +
\tau_{23})=-\frac{\mathrm i}{2}\mathrm e^{\mathrm i (\omega
t_p+\omega_0 \tau_{12}-{\bf k_3r})}\mathrm e^{-\tau_{12}/T_2}\times \\
\nonumber
&&\left[\theta_{3+}(\frac{n_T^{b3}-n_e^{b3}}{2}+J_z^{b3}-S_z^{b3})\cos(\omega_L
\tau_{12}/2)+(\theta_{3-}(J_y^{b3}-\mathrm i
J_x^{b3})-\theta_{3+}(S_y^{b3}-\mathrm i S_x^{b3}))\sin(\omega_L
\tau_{12}/2) \right] \\ \nonumber &&\rho_{24}(3 t_p + 2\tau_{12} +
\tau_{23})=-\frac{\mathrm i}{2}\mathrm e^{\mathrm i (\omega
t_p+\omega_0 \tau_{12}-{\bf k_3r})}\mathrm e^{-\tau_{12}/T_2}\times \\
\nonumber
&&\left[\theta_{3-}(\frac{n_T^{b3}-n_e^{b3}}{2}-J_z^{b3}+S_z^{b3})\cos(\omega_L
\tau_{12}/2)+(-\theta_{3+}(J_y^{b3}+\mathrm i
J_x^{b3})+\theta_{3+}(S_y^{b3}+\mathrm i S_x^{b3}))\sin(\omega_L
\tau_{12}/2) \right] \eea

\subsection{Stimulated photon echo amplitude}

Our analysis shows that the choice of the polarization configuration
for the excitation pulses gives us flexibility for observing
different mechanisms of photon echo generation. There are two
radically different configurations: (i) all pulses are linearly
co-polarized, for example, horizontally (HHH) and (ii) the first and
the second pulses are linearly cross-polarized. For clarity we will
discuss configuration HVV in the following. One can rewrite
Eqs.~\eqref{rho1324} for these two configurations.

\subsubsection{Configuration HHH}

In this case $\theta_{1+}=\theta_{1-} \equiv \theta_{1}$,
$\theta_{2+}=\theta_{2-} \equiv \theta_{2}$, and
$\theta_{3+}=\theta_{3-} \equiv \theta_{3}$. This leads to
$\Delta=\Delta_T=0$, $\Sigma=\Sigma_T=2\theta_{1}\theta_{2}$, and,
therefore, $S_z=S_y=J_z=J_y=0$. The polarization of the stimulated
echo signal is also horizontal: \bea & P_{HHHH} \sim -\mathrm
i\mathrm e^{\mathrm i (\omega t_p+\omega_0
\tau_{12}-{\bf k_3r})}\mathrm e^{-\tau_{12}/T_2}\theta_{3}\times \\
\nonumber &\left[\frac{n_T^{b3}-n_e^{b3}}{2}\cos(\omega_L
\tau_{12}/2)-\mathrm i (J_x^{b3}- S_x^{b3})\sin(\omega_L
\tau_{12}/2) \right] + c.c. \eea From this equation it is obvious, that
only one term of this expression, namely the one which is
proportional to $S_x^{b3}$ decays exponentially with the long
relaxation time of the electron spin, $T_1^e$, when we increase the
delay between the second and the third pulses $\tau_{23}$. The other
terms, which are proportional to $n_T^{b3}$, $n_e^{b3}$ and
$J_x^{b3}$, decay on the shorter time scales $\tau_r$ or $\tau_T$.
One can also clearly see that the long-lived part appears
exclusively in magnetic field.

If we rewrite the previous equation with substitution of all terms,
we obtain: \bea &P_{HHHH} \sim -\frac{\mathrm i}{8}\mathrm
e^{\mathrm i ({\bf k_1-k_2-k_3}){\bf r}}\mathrm
e^{-2\tau_{12}/T_2}\theta_{1}\theta_{2}\theta_{3}\times \\ \nonumber
&\left[\mathrm e^{-\tau_{23}/\tau_r}(2\cos^2(\omega_L
\tau_{12}/2)+\mathrm e^{-\tau_{23}/T_h}\sin^2(\omega_L
\tau_{12}/2))+\mathrm e^{-\tau_{23}/T_1^e}\sin^2(\omega_L
\tau_{12}/2) \right] +c.c.\label{phhhh} \eea The long-lived part is
expected to show up as a constant background at fixed magnetic
field, when the stimulated echo is measured as function of
$\tau_{23}$, or as slowly oscillating background when the echo is
measured as function of magnetic field.

\subsubsection{Configuration HVV}

In this configuration $\theta_{1+}=\theta_{1-} \equiv \theta_{1}$,
$\theta_{2+}=-\theta_{2-} \equiv \theta_{2}$,
$\theta_{3+}=-\theta_{3-} \equiv \theta_{3}$.
This leads to $\Delta=-\Delta_T=2\theta_{1}\theta_{2}$ and  $\Sigma=\Sigma_T=0$.

The polarization of the stimulated echo signal is horizontal:
\bea &P_{HVVH} \sim -\mathrm i\mathrm e^{\mathrm i (\omega
t_p+\omega_0
\tau_{12}-{\bf k_3r})}\mathrm e^{-\tau_{12}/T_2}\theta_{3}\times \\
\nonumber &\left[(J_z^{b3}- S_z^{b3})\cos(\omega_L \tau_{12}/2)-
(J_y^{b3}+ S_y^{b3})\sin(\omega_L \tau_{12}/2) \right] + c.c.\eea Clearly
this signal contains long-lived and short-lived parts. After
substitution of $S_{z,y}^{b3}$ and $J_{z,y}^{b3}$, we obtain: \bea
&P_{HVVH} \sim -\frac{\mathrm i}{8}\mathrm e^{\mathrm i
({\bf k_1-k_2-k_3}){\bf r}}\mathrm
e^{-2\tau_{12}/T_2}\theta_{1}\theta_{2}\theta_{3}\times \\ \nonumber
&\left[\mathrm e^{-\tau_{23}/\tau_T}(\cos(\omega_L
\tau_{12})+\cos(\omega_L \tau_{12}/2)[\xi_1\cos(\omega_L
\tau_{12}/2)+\xi_2\sin(\omega_L \tau_{12}/2))] \right. \\ \nonumber
&\left. +\mathrm e^{-\tau_{23}/T_2^e}\left[\cos(\omega_L
(\tau_{12}+\tau_{23}))- \cos(\omega_L \tau_{12}/2)[\xi_1\cos(\omega_L
(\tau_{12}/2+\tau_{23}))+\xi_2\sin(\omega_L
(\tau_{12}/2+\tau_{23}))]\right] \right] \\ \nonumber &
\quad \quad\quad \quad\quad \quad\quad \quad\quad \quad\quad \quad\quad \quad\quad \quad\quad \quad\quad \quad\quad \quad\quad \quad +c.c.\eea If trion spin relaxes
before trion recombination, $T_h \ll \tau_r$, then $\xi_1=\xi_2=0$
and $P_{HVVH}$: \bea &P_{HVVH} \sim -\frac{\mathrm i}{8}\mathrm
e^{\mathrm i ({\bf k_1-k_2-k_3}){\bf r}}\mathrm
e^{-2\tau_{12}/T_2}\theta_{1}\theta_{2}\theta_{3}\times \\ \nonumber
&\left[\mathrm e^{-\tau_{23}/\tau_T}\cos(\omega_L \tau_{12})
+\mathrm e^{-\tau_{23}/T_2^e}\cos(\omega_L
(\tau_{12}+\tau_{23}))\right] + c.c.\eea One sees from this equation, that
the amplitude of the long-lived signal at zero magnetic field and
the amplitude of the long-lived oscillations in magnetic field are
the same. If we sum $P_{HVVH}$ over all $\omega_L$ to take into
account a possible spread of Larmor frequencies, then the
oscillating signal in magnetic field can decay faster than the
signal in zero magnetic field.

If the trion spin relaxes slowly, $T_h \gg \tau_r$, then the
amplitude of the long-lived signal depends strongly on magnetic
field. In zero magnetic field $P_{HVVH}$ is: \bea P_{HVVH} \sim
-\frac{\mathrm i}{8}\mathrm e^{\mathrm i ({\bf k_1-k_2-k_3}){\bf r}}\mathrm
e^{-2\tau_{12}/T_2}\theta_{1}\theta_{2}\theta_{3}\times
\left[\mathrm e^{-\tau_{23}/\tau_T}(1+\xi_1) \right. \left. +\mathrm
e^{-\tau_{23}/T_2^e}(1-\xi_1) \right] + c.c.\eea

The long-lived part of the signal is proportional to $1-\xi_1
\approx \tau_r/(T_h+\tau_r)$, therefore if $T_h \gg \tau_r$, then
the signal at $B=0$ vanishes. This is due to the fact that after the
second pulse a long-lived spin polarization $S$, which is
responsible for the long-lived echo signal, does not appear: after
trion recombination the spin state of the resident carrier does not
differ from that before trion formation.

In a transverse magnetic field the change of the relative
orientation of the electron spin and the trion spin leads to
appearance of spin polarization of the resident electrons, even if
the spins of the carriers in the trions do not relax
\cite{rsa_vs_ml}. With increasing magnetic field this imbalance
leads to an increase of the stimulated photon echo amplitude.

\subsubsection{Stimulated photon echo before the first pulse arrival}

It is worth recalling that all pulses in our experiment arrive
periodically with the laser repetition period $T_R \approx 13$~ns.
The stimulated photon echo signal before the first pulse arrival
(see Fig.~3) is in fact the signal created by the preceding pair of
pulses 1 and 2. At $\tau_{23}$ close to $T_R$ changes of the initial
conditions for $\rho$ by the third pulse before the arrival of the
first pulse have to be taken into account. For the configurations
with linearly polarized pulses we have to consider only changes of
$n_e$ and $n_T$ induced by the previous three pulses, because the
other components of spin or polarization contain phase factors and
do not contribute to the signal. If $\tau_{23}$ close to $T_R$,
$n_e$ and $n_T$ are changed only by the third pulse then. As a
result, the density matrix before the subsequent first pulse arrival
is given by:
\begin{align}
\rho^{b1} &=\frac{1}{2}
\begin{pmatrix}
n_e^{b1} & 0  &0   &0   \\
0 &n_e^{b1}& 0&0\\
0&0&n_T^{b1}&0\\
0&0&0&n_T^{b1}
\end{pmatrix}.
\end{align}
Here $n_T^{b1}=\theta_3^2\mathrm e^{-\tau_{31}/\tau_r}/4$
$n_e^{b1}=1-\theta_3^2\mathrm e^{-\tau_{31}/\tau_r}/4$, $\tau_{31}$
is the temporal separation between the third pulse and the next
first pulse, $\tau_{31}=T_R-(\tau_{12}+\tau_{23})$.

These initial conditions lead for the signals $P_{HHHH}$ and
$P_{HVVH}$ only to one additional factor $\sim
(n_e^{b1}-n_T^{b1})=1-\theta_3^2\mathrm e^{-\tau_{31}/\tau_r}/2$.
This factor corresponds to an exponential decrease of the echo
signal at negative delay $\tau_{23}$ close to the first pulse. In
our case this decrease can be seen in $P_{HHHH}$ through the
long-lived component of the signal in magnetic field, which still
exists in the relevant temporal range as it does not decay due to a
Larmor frequency spread so that it is preserved until the next first
pulse arrival.

\clearpage
\begin{figure}
\begin{center}
\includegraphics[width=10cm]{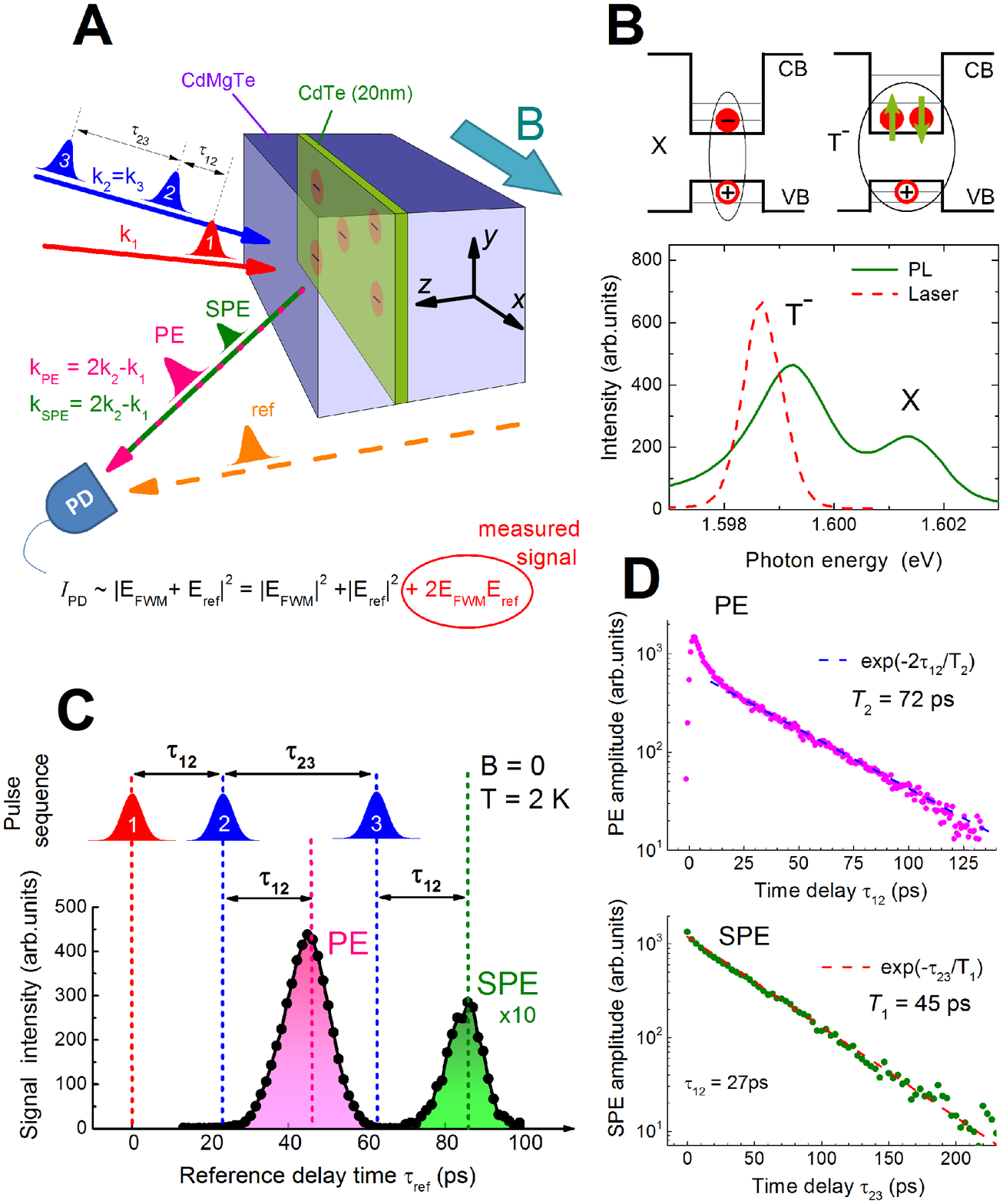}
\end{center}
\end{figure}

\noindent {\bf Fig. 1. Scheme of photon echo experiment and optical properties of investigated structure.} (A) The CdTe/(Cd,Mg)Te quantum well (QW) is optically excited with a sequence of three laser pulses with variable delays $\tau_{12}$ and $\tau_{23}$ relative to each other. The resulting four-wave mixing transients $|E_{\rm FWM}(t)|$ are detected in $2\bf{k_2-k_1}$ direction using heterodyne detection. All measurements are performed at temperature of 2~K. (B) Top: schematic presentation of exciton ($X$) and trion ($T^-$) complexes in QW. The QW potential of conduction (CB) and valence (VB) bands leads to spatial trapping of electrons and holes. Bottom: Photoluminescence (PL) spectrum (solid line) measured for above-barrier excitation with photon energy 2.33~eV, demonstrating $X$ and $T^-$ emission. The laser spectrum (dashed line) used in photon echo experiment is tuned to the low energy flank of $T^-$ emission line. (C) Four-wave mixing transients for $\tau_{12}=23$~ps and $\tau_{23}=39$~ps. Spontaneous (PE) and stimulated (SPE) photon echo signals appear at $\tau_{\rm ref}=2\tau_{12}$ and $\tau_{\rm ref}=2\tau_{12}+\tau_{23}$, respectively. (D) Decay of PE and SPE peak amplitudes. From exponential fits (dashed lines) we evaluate $T_2=72$~ps and $T_1=45$~ps.

\clearpage
\begin{figure}
\begin{center}
\includegraphics[width=17cm]{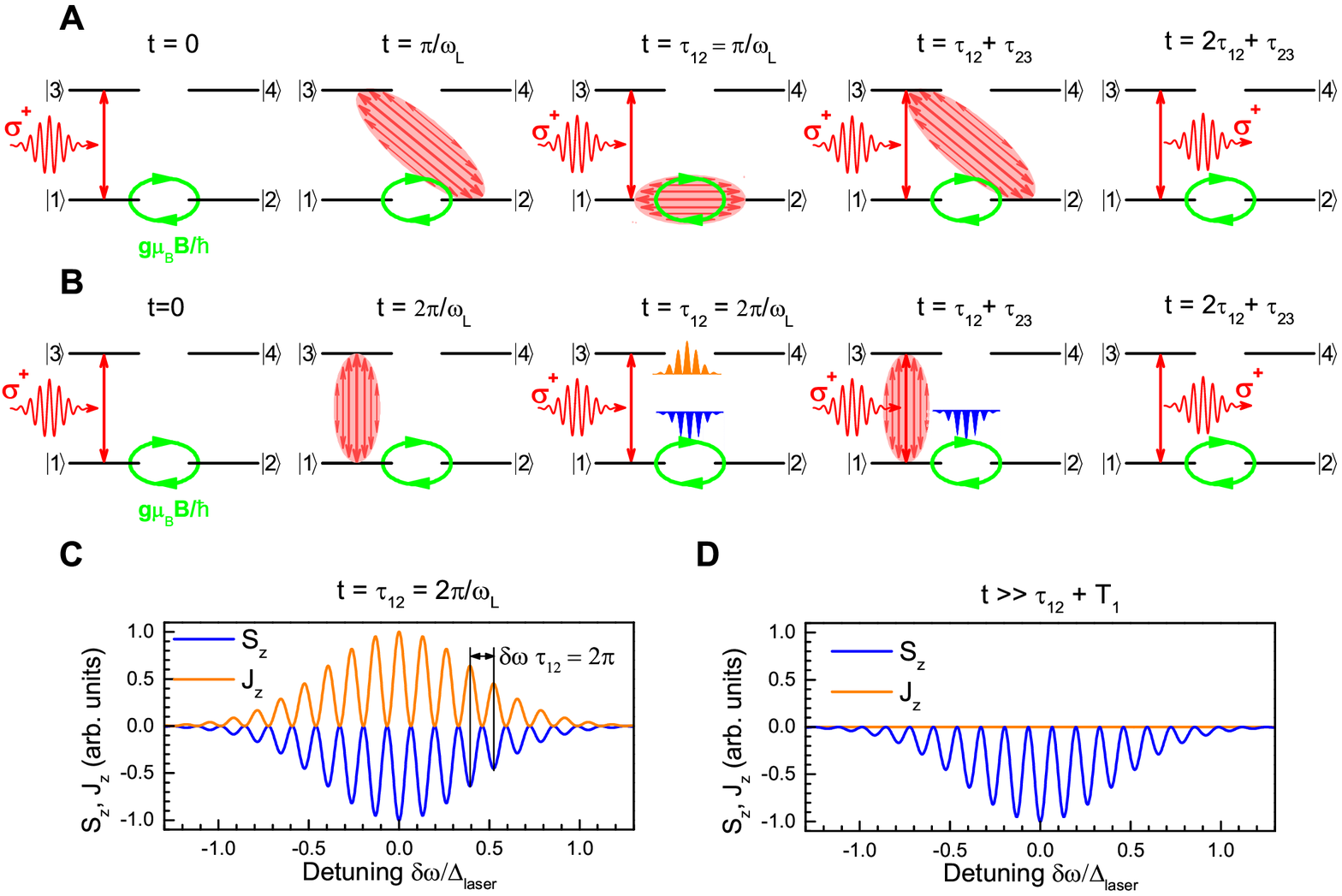}
\end{center}
\end{figure}

\vskip 5mm

\noindent {\bf Fig. 2. Schematic presentation of the main mechanisms responsible for magnetic-field-induced stimulated photon echoes (SPE).} Optical pulses are circularly polarized. (A) Transfer of optical coherence into electron spin coherence ($S_x$ and $S_y$ components). The efficiency is maximum for $\tau_{12}=\pi/\omega_L$. (B) Creation of spectral spin fringes for electrons and trions ($S_z$ and $J_z$ components). This mechanism is most efficient for $\tau_{12}=2\pi/\omega_L$. The spectral spin gratings for electrons and trions are shown in (C) at the moment of creation by the second pulse ($t=\tau_{12}=2\pi/\omega_L$) and in (D) after trion recombination and before arrival of pulse 3 ($t\gg\tau_{12}+T_1$).

\clearpage
\begin{figure}
\begin{center}
\includegraphics[width=11cm]{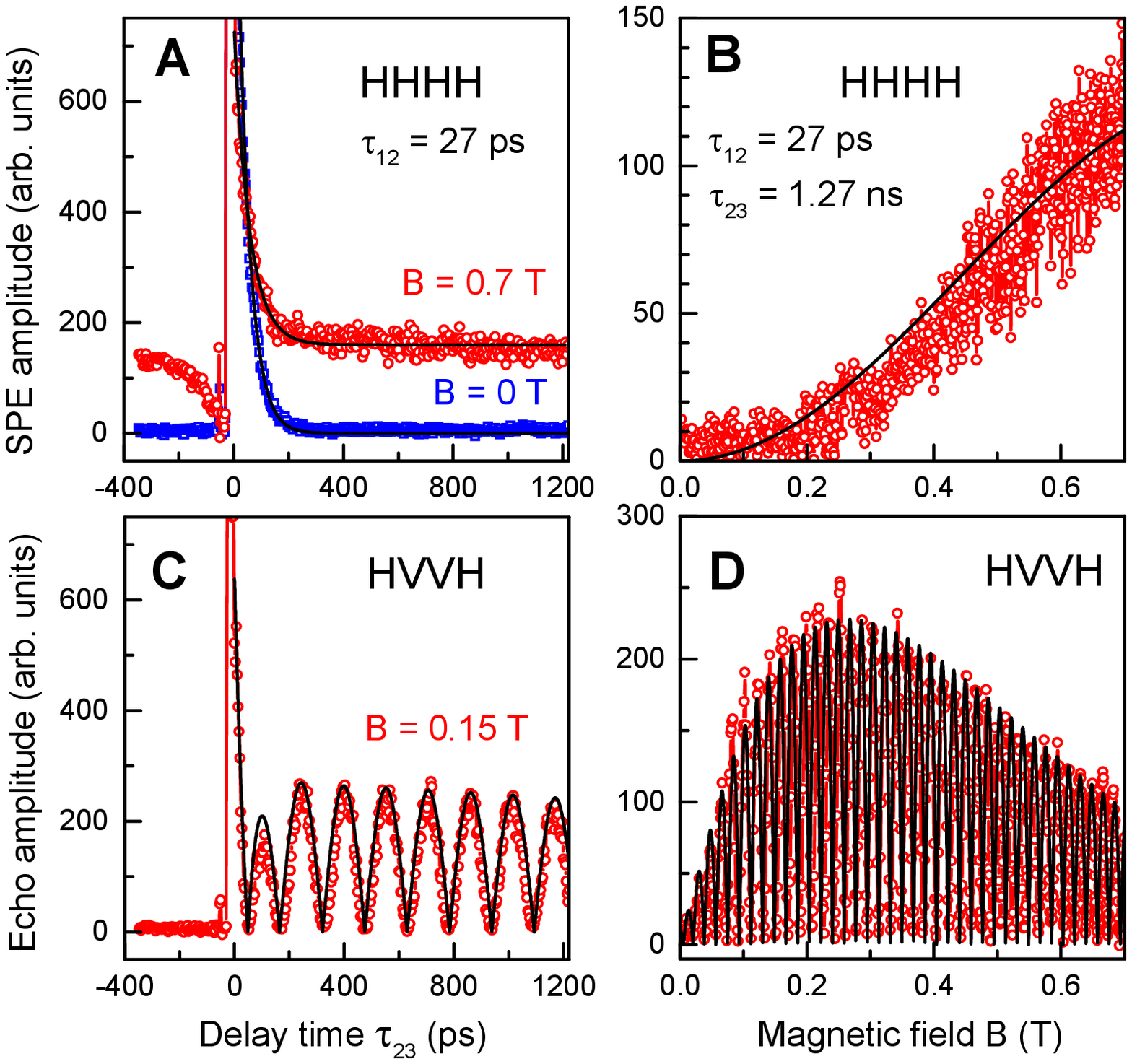}
\end{center}
\end{figure}

\vskip 5mm

\noindent {\bf Fig. 3. Experimental demonstration of magnetic-field-induced long-lived stimulated photon echo (SPE).} The delay time $\tau_{12}$  is set to 27~ps. (A) Stimulated echo amplitude in configuration HHHH as function of time delay $\tau_{23}$ at $B=0$ and 0.7~T. (B) Long-lived part of the stimulated echo amplitude as function of magnetic field. Dots are experimental data. Black curves in A and B give theoretical calculations with parameters: $g=1.52$, $\tau_r=43$~ps, $T_h=1000$~ps, and $T_1^e \sim 50$~ns. (C) Stimulated echo amplitude in configuration HVVH as function of the delay $\tau_{23}$ at $B=0.15$~T. (D) Long-lived part of stimulated echo amplitude versus magnetic field. Black curves in C and D are calculations with parameters: $g=1.52$, $\Delta g = 0.018$, $\tau_r=43$~ps, $T_h=1000$~ps, and $T_2^e > T_R$.


\begin{thebibliography}{1}
\bibitem{fnt1} Also localization of an electron at a donor located in QW is possible. In this case optical transitions between donor bound electron and donor bound exciton
($D^0X$) are addressed. Most importantly, however, the same SPE scenarios are valid for the trion and donor-bound exciton complexes.
\bibitem{fnt2} Note that the density of electrons in the QW is low (exciton Bohr radius $a_B \ll \sqrt{n_e}$), which allows us to consider the resident electrons as isolated,
non-interacting carriers. Many-body interactions are weak and can be neglected at the low excitation densities used in our experiment (pulse energies below 100~nJ cm$^{-2}$).
\bibitem{fnt3} For the neutral exciton the transitions at zero magnetic field are not circularly but linearly polarized. Moreover, in magnetic field the exciton level splitting is
determined by electron and hole Zeeman splitting, and not only by the hole one.
\end{thebibliography}
\end{document}